# Analysis of nonlinear conductivity of point contacts on the base of FeSe in the normal and superconducting state.

Yu.G. Naidyuk[1], N.V. Gamayunova[1], O.E. Kvitnitskaya[1], G. Fuchs[2], D.A. Chareev[3], A.N. Vasiliev[4,5,6]

[1]*B. Verkin Institute for Low Temperature Physics and Engineering, National Academy of Sciences of Ukraine, 47 Lenin Ave., 61103 Kharkiv, Ukraine*

[2]*Leibniz-Institut für Festkörper- und Werkstoffforschung Dresden e.V., Postfach 270116, D-01171 Dresden, Germany*

[3]*Institute of Experimental Mineralogy, Russian Academy of Sciences, 142432 Chernogolovka, Moscow District, Russia*

[4]*Low Temperature Physics and Superconductivity Department, Physics Faculty, M.V. Lomonosov Moscow State University, 119991 Moscow, Russia*

[5]*Theoretical Physics and Applied Mathematics Department, Ural Federal University, 620002 Ekaterinburg, Russia*

[6]*National University of Science and Technology "MISiS", Moscow 119049, Russia*

## Abstract

Nonlinear conductivity of point contacts (PCs) on the base of FeSe single crystals has been investigated. Measured *dV/dI* dependencies demonstrate the prevailing contribution to the PC conductivity caused by the degraded surface. Superconducting (SC) feature in *dV/dI* like a sharp zero-bias minimum develops for relatively low ohmic PCs, where the deep areas of FeSe are involved. Analysis of *dV/dI* has shown that the origin of the zero-bias minimum is connected with the Maxwell part of the PC resistance, what masks energy dependent spectral peculiarities. Even so, we have found the specific features in *dV/dI* – the sharp side maxima, which may have connection to the SC gap, since their position follows the BCS temperature dependence. Exploring the *dV/dI* spectra of the rare occurrence with Andreev-like structure, the two gaps with $\Delta$=2.5 and 3.5 meV were identified.

## Introduction

FeSe compound, belonging to the 11- structure groups of iron based superconductors, is actively investigated nowadays. On one hand, this is due to the fact that FeSe has the simplest crystal structure among other superconducting iron chalcogenides and pnictides. Besides, it shows only the structural phase transition at $T_S \sim 100K$, without an accompanying magnetic phase transition. On the other hand, the superconducting (SC) transition temperature $T_c \sim 9K$ [1] in FeSe increases drastically under pressure up to 37K [2] and $T_c$ reaches incredible 100K in the case of a FeSe monolayer [3].

Observation of Shubnikov–de Haas oscillations demonstrates the low carrier density (~0.01 carriers/Fe) and the small Fermi energy (~3.6 meV). The Fermi surface occupies only a small part of the Brillouin zone and contains probably one electron and one hole thin cylinder [4]. The electronic structure of the low-temperature orthogonal FeSe-phase is similar to that for almost compensated semimetals with ultrafast electron-like minority carriers having small density of about $10^{18}$ cm$^3$ [5]. These carriers may occur during formation of a "Dirac cone" or in the case of the significant anisotropy of the Fermi surface.

Tunnel $dI/dV$ spectra of FeSe demonstrate a V-shaped zero-bias minimum with side maxima at +/-2.5 meV and shoulders at +/-3.5 meV, which were taken as the evidence for the two-gap SC state [6]. Thus, the Fermi energy $E_F$ in FeSe is comparable to the value of the SC gap(s) $\Delta$: $\Delta/E_F \sim 1$ (~ 0.3) for the electron (hole) band [6]. As a result, the BCS (Bardeen–Cooper–Schriffer)–BEC (Bose-Einstein condensation) crossover in FeSe can be realized.

All mentioned features make FeSe very attractive for point-contact (PC) investigations [7]. This work presents the study of current-voltage $I(V)$ characteristics and their derivatives $dV/dI(V)$ of PCs based on FeSe single crystals ($T_c = 9K$) [1] in the normal and SC state. PC measurements of nonlinear $I(V)$ curves and their derivatives are used in the method of Yanson PC spectroscopy [7] to identify specific bosonic (phononic) excitations and to obtain information about the SC gap utilizing PC Andreev-reflection spectroscopy.

## Results

The plate-like single crystals of FeSe$_{1-x}$ (x=0.04 +/-0.02) superconductor were grown in evacuated quartz ampoules using flux technique as described in [1]. The resistivity and magnetization measurements revealed a SC transition temperature up to $T_c = 9.4$ K. PCs were established by touching of a sharpened thin Cu wire (or Ag and W wires) to cleaved by a scalpel at room temperature flat surface of FeSe or contacting by the wire an edge of plate-like samples. Thus, we have measured heterocontacts between normal metal and the title compound. The differential resistance $dV/dI(V) \equiv R(V)$ of PC were recorded by sweeping the dc current $I$ on which a small ac current $i$ was superimposed using a standard lock-in technique. The measurements were performed in the temperature range from 3 K to slightly above $T_c$. No principal difference was found by "flat" or "edge" PC geometry in $dV/dI(V)$ data, because $dV/dI(V)$ variate more significantly from one PC to another. Several attempts have been made with FeSe surface prepared by polishing using very soft sand paper or even office paper, but it was more difficult to obtain the SC features in the PC spectra in the latter case.

As shown in Fig.1, the $dV/dI$ spectra of PCs demonstrate overall "semiconducting" behavior (the negative $dV/dI$ curvature) representing a broad maximum centered at zero-bias voltage, which is more pronounced with increasing of the PC resistance. For decreasing PC resistance, the measured below $T_c$ $dV/dI$ spectra tend to have a «V»-shaped sharp zero-bias minimum (see Fig.1).

Fig.2 shows $dV/dI$ for two PCs demonstrating "semiconducting" and "metallic" behavior with the sharp zero-bias minimum developing below $T_c$ both for "semiconducting" and "metallic" behavior. Note, that in spite of the different "semiconducting" and "metallic" shape of $dV/dI$, both of them show a similar asymmetry (see right inset of Fig. 2). Fig.3 displays $dV/dI$ with the "metallic" behavior and a much sharper zero-bias dip compared to those in Fig. 2. In this case $dV/dI$ above $T_c$ shows a shallow zero-bias maximum. A more complicated *dV/dI* shape develops for PC in Fig.4, where the zero-bias minimum pattern is more complex with additional sharp side peaks. It turned out, that the position of the main side peak follows the BCS temperature dependence.

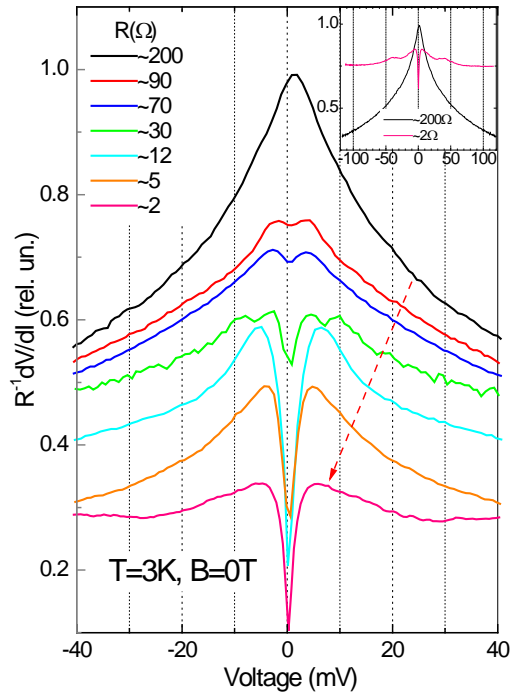

Fig.1. Series of *dV/dI* curves at decreasing of PC resistance from about 200Ω to 2Ω (from the upper curve to the bottom one). The curves, except the upper one, are shifted down for clarity. Pronounced zero-bias minimum develops along with the transition from "semiconducting" (high resistance) to more "metallic" (low resistance) behavior of *dV/dI*. Inset shows *dV/dI* for two PCs from the main panel at larger bias.

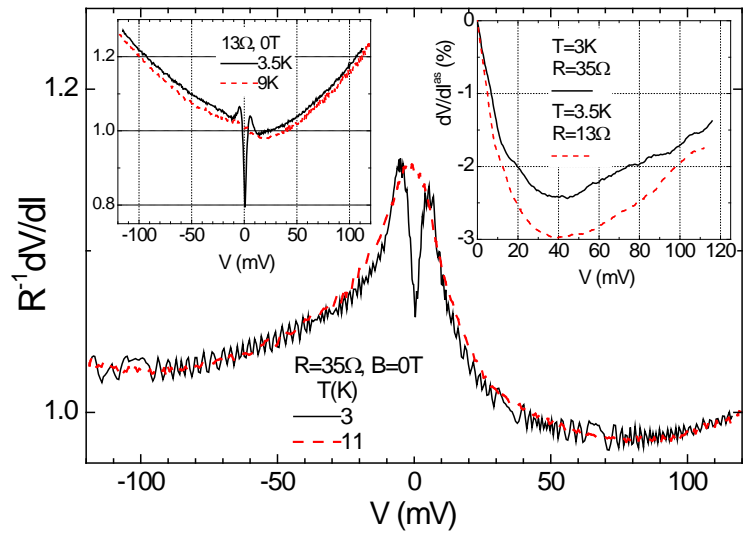

Fig.2. Typical *dV/dI* spectra (the main panel and left inset) of two FeSe-Cu PCs measured well below and just above $T_c$. Right inset shows the antisymmetric part $dV/dI^{as}(\%) = 100[dV/dI\,(V>0) - dV/dI\,(V<0)]/2dV/dI\,(V=0)$ of *dV/dI* calculated for both contacts at low temperatures.

A rarely observed *dV/dI* is shown in Fig. 5. Here, *dV/dI* measured at the low temperature of 4.2K demonstrates the Andreev-like double minimum structure around zero-bias. The position of the minima is about +/-3.5mV, what is close to the large gap value (3.5 mV) in FeSe measured by tunneling spectroscopy in [6].

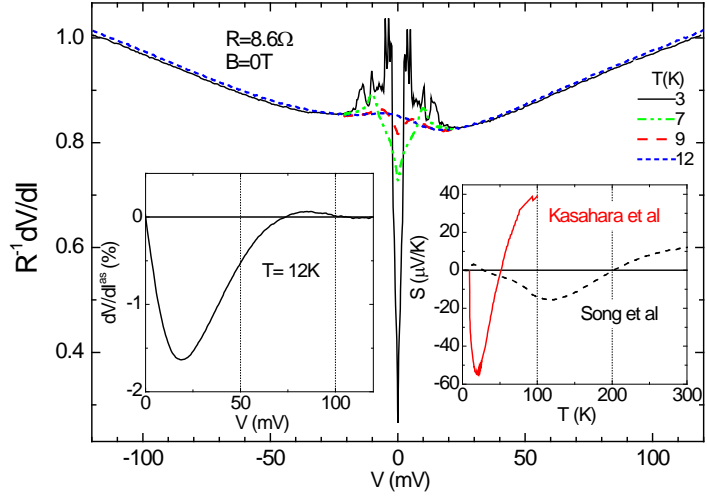

Fig.3. Temperature variation of the *dV/dI* spectrum (main panel) of FeSe-Cu PC. Left inset shows the antisymmetric part $dV/dI^{as}(\%) = 100[dV/dI\ (V>0) - dV/dI\ (V<0)]/2dV/dI\ (V=0)$ of *dV/dI* calculated for *dV/dI* at T=12K. Right inset shows the behavior of thermo-emf in single FeSe crystals according to Kasahara et al. [6] and Song et al [8].

**Discussion**

"Semiconducting" behavior of *dV/dI* can be due to the low concentration of carriers and/or depleted (semiconducting) surface layer. As many investigations show, the transport properties of FeSe are very sensitive to the stoichiometry and the distribution of Fe vacancies. Thus, Chen et al. [9] reported about the observation of three different Fe-vacancy orders and among them one was identified to be nonsuperconducting and magnetic at low temperature. Also Chang et al. [10] discussed the amorphous oxide on the surface of the fresh FeSe nanowires, which becomes thicker with prolonged air exposure. Greenfield et al. [11] underlined that "Vacancies in the iron sublattice and the incorporation of disordered oxygen-containing species are typical for nonsuperconducting antiferromagnetic samples, whereas a pristine structure is required to preserve superconductivity. Exposure to ambient atmosphere resulted in the conversion of superconducting samples to antiferromagnetic ones". Therefore, we believe that the "semiconducting" *dV/dI* shape is due to the degraded on air thick surface layer. By decreasing of the PC resistance, we "penetrate" deeper to the bulk material. As a result, *dV/dI* becomes more "metallic" and the SC zero-bias minimum developes.

Interestingly, in the recent report by Ventzmer et al. [12], they measured similar "semiconducting" type of *dI/dV* in the planar tunneling junctions FeSe/AlOx/Ag patterned lithographically into mesastructures. They observed also a metallic like behavior in PC noticing that a tunneling barrier with pinholes can result in a large variety of structures in the differential conductivity. Sooth to say, *dI/dV* characteristics in [12] resemble a little the tunneling behavior, since their variation with a bias was less than a factor of two and for some PCs only a few percent.

The lack of characteristic Andreev reflection features in the *dV/dI* spectra below T$_c$ (like double minima structure instead of sharp zero-bias minimum) can be related to the realization of the thermal regime [7,13] of the current flow in PC. This regime develops in materials with high resistivity, where inelastic mean free path becomes smaller that the PC size (diameter) *d*. In this case, the resistivity ρ(T) determines the behavior of *I(V)* characteristics and their *dV/dI* derivatives according to the equation [7,13]:

$$I(V) = Vd \int_0^1 \frac{dx}{\rho(T_{PC}(1-x^2)^{1/2})}, \qquad (1)$$

while the temperature in the PC core $T_{PC}$ increases with a voltage *V* according to the relation

$$T_{PC}^2 = T_0^2 + V^2/4L_0, \qquad (2)$$

where $T_0$ is a bath temperature, $L_0 = 2.45 \cdot 10^{-8}$ V$^2$/K$^2$ is the standard Lorentz number. In the case of $T_{PC} \gg T_0$, the temperature in the PC core $T_{PC}$ increases linearly with the applied voltage $T_{PC} = V/2\sqrt{L_0}$ with the rate 3.2K/mV.

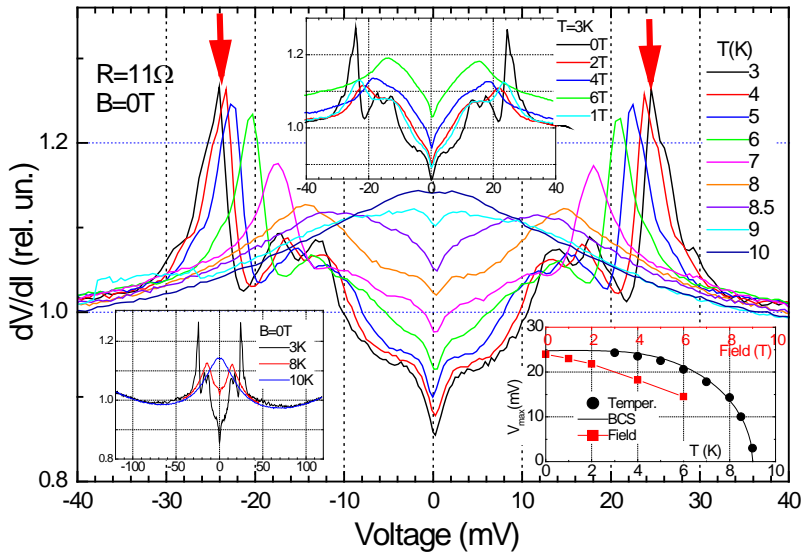

Fig.4. Temperature variation of the *dV/dI* spectrum of FeSe-Cu PC with the pronounced side peaks. Upper inset: *dV/dI* for the same contact in magnetic field at T=3K. Left inset shows *dV/dI* at a few temperatures at larger bias. Right inset shows the temperature and magnetic field position of the side peak.

By fitting Eqs. (1) and (2) to the measured *dV/dI(V)* (see Fig.6), the following parameters were estimated: the Lorentz number in FeSe L≈9L$_0$, the PC residual resistivity $\rho_0$ ≈ 0.35 mΩ·cm, the PC diameter d ≈ 0.8 μm for the PC resistance of about 5 Ω. The obtained large value of 9L$_0$ for the Lorentz number in FeSe correlates with its estimation from the thermal conductivity and resistivity data just above T$_c$ at 10K in [6]. The rather large of $\rho_0$ can be attributed to the degraded surface and other imperfections on the surface in the contact area.

The asymmetry of the *dV/dI* characteristics in the thermal regime in the case of heterocontacts is connected with thermo-emf, so that antisymmetric part of *dV/dI* is proportional to the difference between the Seebeck coefficients *S(T)* of the contacting metals [14,15]. As shown in the insets in Figs. 2 and 3, *dV/dI*$^{as}$ looks qualitatively similar to the temperature dependence of *S(T)* in FeSe indicating that the PCs are in the thermal regime. Note, that in spite of different "metallic" and "semiconducting" shape of *dV/dI* in Fig.2, their antisymmetric parts are similar. That is the antisymmetric part of *dV/dI* is more reproducible and reflects rather the massive (bulk) material properties (see also Appedix B in [16] for the discussion). Here, it should be mentioned that the behavior of *S(T)* in FeSe samples measured by different authors is different (see, e.g., the inset in Fig.3). It is known that the thermo-emf is the most sensitive transport property of metals: it is some kind of derivative of conductivity and it depends strongly on the electronic structure [17]. Because of that, the Seebeck coefficient is very sensitive to the quality of FeSe samples, much more than the resistivity.

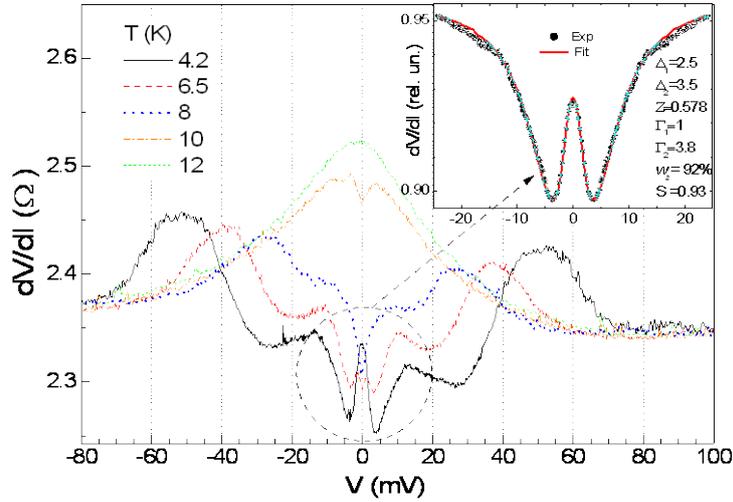

Fig.5. Temperature variation of the *dV/dI* spectrum of FeSe-Ag PC with Andreev-like double minimum at zero bias and lowest temperature. Inset shows a fit (solid red curve) of the normalized on the normal state *dV/dI* at 4.2K (symbols) using the two gap model [26] with the parameters shown in the panel. Here, $\Delta$ and $\Gamma$ are in meV. S is the scaling factor, which reflects the difference in intensities of experimental and calculated curves. In the ideal case it must be S=1. *w* is the partial contribution of the larger gap 3.5meV to the calculated spectra.

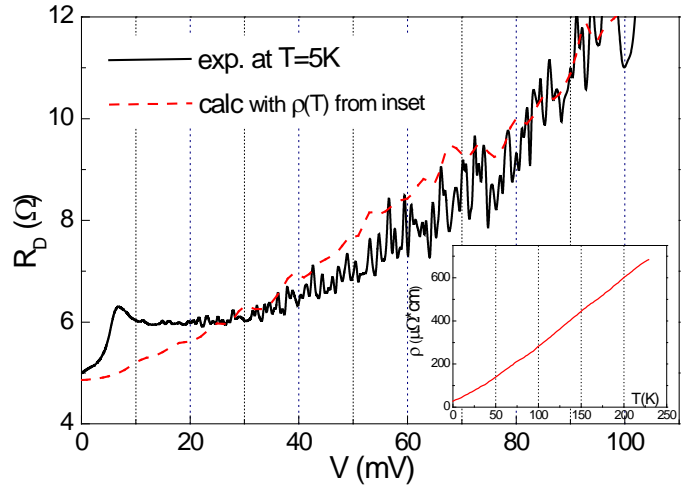

Fig. 6. Fit (dashed red curve) of the *dV/dI* spectrum (solid black curve) of FeSe-W PC above the SC minimum (>20mV) according to Eqs. (1, 2) with the parameters $d\approx 0.8\mu m$, $\rho_0\approx 0.35$ m$\Omega$cm and $L=9L_0$. Inset shows the resistivity $\rho(T)$ of FeTe single crystal adapted from [1] and used in Eq. (1), which is additionally increased by an amount of the enhanced residual resistivity $\rho_0$ in PC.

Let us turn to the discussion of the origin of the sharp zero-bias minimum. Obviously, it is connected with the SC state in PC. At the same time, the nature of this SC dip has to be clarified. Such zero-bias dip in *dV/dI* (maximum in *dI/dV*) is often connected with the Andreev bound states in the case of unconventional d-wave superconductors. However, a similar structure is observed regularly in simple elemental (conventional) superconductors [18]. Especially, such dip in *dV/dI* is characteristic for the complex SC compounds with high residual resistivity like high-Tc materials

[19], heavy-fermion systems [20] and amorphous alloys [21]. Gloos et al. [20] concluded that such zero-bias dip is due to the Maxwell's resistance (see Eq. (3)) being suppressed in the SC state.

Let us try to estimate parameters of PC from its resistance $R_{PC}$. The latter is expressed by the well known Wexler formula, which contains the sum of ballistic Sharvin and diffusive Maxwell resistance:

$$R_{PC} \approx 16\rho l/3\pi d^2 + \rho/2d, \qquad (3)$$

where $\rho l = p_F/ne^2 \approx 1.3 \times 10^4 n^{-2/3} \approx 3.2 \times 10^{-10}$ $\Omega$ cm$^2$, using the carrier density n$\approx 2.53 \times 10^{20}$ cm$^{-3}$ from [22]. The residual resistivity $\rho_0$ in the PC core is unknown in Eq.(3). If we suppose that $\rho_0 \approx 0.035$ m$\Omega$ cm just above $T_c$ like in the bulk FeSe crystal [1], then, according to Eq.(3), a PC diameter of d $\approx 120$ nm and an electron mean free path of l$\approx 90$ nm are estimated for the PC with the resistance of about 5 $\Omega$. That is, d $\approx$ l and the current regime in the investigated PC is neither ballistic, nor diffusive. Moreover, such PC is affected by a high current density j$\approx$V/Rd$^2$, increasing with the rate of about $1.4 \times 10^6$ A/cm$^2$ per 1 mV. On the other hand, the corresponding parameters estimated by fitting of the experimental *dV/dI* curve with similar resistance in Fig. 6 by Eqs. (1) and (2) are d$\approx$ 0.8 $\mu$m and $\rho_0 \approx 0.35$ m$\Omega$ cm. That is, $\rho_0$ is one order of magnitude larger than that in the bulk. Correspondingly, *l* is ten times smaller and this PC is in the diffusive limit d>>*l*. This is due to a degraded surface layer resulting in a higher resistivity compared to the bulk. If we take the last calculated parameters for that PC and use Eq.(3), then the Maxwell contribution to the PC resistance estimated from Eq. (3) exceeds the ballistic Sharvin resistance by more than one order of magnitude. Also the current density in this case will be still high, i.e. it increases with the rate about $3 \times 10^4$ A/cm$^2$ per 1 mV [1]. Thus, as Gloos et al. concluded [20], the resistive Maxwell term contributes mainly to the observed SC sharp minimum. Recovering the Maxwell resistance, which is zero in the SC state, due to increasing of the current density and/or temperature in the PC core in consequence of Joule heating produces a zero-bias minimum. Because of the coherence length in FeSe (equal 1.3 and 5.7 nm for the *c* and *ab* directions, respectively [4]) is also much smaller than the PC size (diameter), the transition of the PC core in the normal state due to increasing current density will occur smoothly involving successively further (deeper) regions.

Let's consider the sharp side peaks shown in Fig.4. Their temperature behavior corresponds well to the BCS curve. So, it looks like this feature is somehow connected with the SC order parameter or gap. Sharp peaks in *dV/dI* may be connected with the abrupt transition from SC to the normal state of some region in PC. To result in such sharp transition, this region must be smaller than the coherence length, which is less than 5.7 nm [4]. More likely, we have a multicontact structure in this case with at least one PC with the size of the order or less than the coherence length[2]. For such small PC the suppression of superconductivity may occur due to reaching of pair-breaking current density j$\approx$en$\Delta/p_F \approx$en$^{2/3}\Delta/3\hbar$ [25]. Using n$\approx 2.53 \times 10^{20}$ cm$^{-3}$ from [22], we get j$\approx 10^7$ $\Delta$ [mV] A/cm$^2$, where $\Delta$ is in mV units. On the other hand, PC with such small dimension (below the coherence length) is in the ballistic limit, where current density depends only on the applied bias j=V/R$_{sh}$d$^2$=V/(16$\rho l/3\pi d^2$) d$^2 \approx$V/$\rho l$, so that j$\approx 3 \times 10^6$V[mV] A/cm$^2$, where V is in mV units. Thereby, current density in such PC reaches the above estimated pair-breaking current density for $\Delta$=2-3 mV at V=7-10 mV, what is not far from the side peak position in Fig.4 taking into account our rough estimation. In this way, assuming that the side peaks are due to reaching of pair-breaking current density and therefore that they are connected to the SC gap value, we can explain the BCS-like dependence of the SC gap in FeSe (or some averaged gap because of the multiband FeSe electronic structure).

Let us return to the spectrum with the Andreev-like double minimum in Fig.5. This structure transforms at first in a zero-bias minimum and then vanishes above 8K, which is close to $T_c$ of the bulk sample. Such transformation of the double minimum is due to the movement of broad side maxima, which position shifts to zero voltage with increasing temperature. So, in our opinion, the conductivity of this PC is governed by two contributions: Andreev reflection and Maxwell term (resistance), which was discussed above. The fitting [3] of the AR structure by the two-gap model [26] results in the gap values Δ=2.5 and 3.5 meV, with the about 90% contribution to the conductivity coming from the large gap. These values are the same as the resolved ones from the tunneling spectra in [6]. It is clear, that some variation of extracted data using seven fitting parameters is possible, however the gap(s) value(s) must concentrate(s) around the minima position of about 3.5meV in any case. Extracted gaps values result in large $2\Delta/k_BT_c$ ratios from 6 to 8, testifying strong coupling superconductivity in FeSe.

---
[1] Note, that the critical current density measured for thin epitaxial films [23] and single crystals [24] in FeSe is of the order of $10^4$A/cm$^2$.

[2] Several of sharp side peaks in Fig. 4 testify about a couple of such PCs.

## Conclusion

We have investigated nonlinear conductivity of PCs on the base of FeSe single crystals. Degraded surface layer (due to oxidation, apparently deviation from stoichiometry and perhaps disturbed through the mechanical PC creation) vastly contributes to the nonlinear conductivity resulting regularly in its non-metallic behavior. This prevents largely to get spectroscopic information from more bulky material. SC features in *dV/dI* develop mainly due to resistive (Maxwell) term in the PC resistance because of failure of ballistic regime in PC. We estimated some material parameters in PC and found that PC has an order of magnitude larger residual resistivity than the bulk material. Also the estimated Lorentz number is strongly enhanced, but this is in line with the results of measurements of thermal and electronic conductivity of FeSe single crystal. Probably, creation of the PC "in situ" on a cleaved surface at ultra high vacuum and low temperatures will help to get rid of degraded surface layer and receive more detailed spectroscopic information. Still, as a practical result, we have found specific features in *dV/dI*, which have connection to the SC gap and allow us to monitor its BCS temperature dependence. As well as, exploring the *dV/dI* spectra of the rare occurrence with Andreev-like structure, the two gaps with Δ=2.5 and 3.5 meV were retrieved.

## Acknowledgments

Funding by the National Academy of Sciences of Ukraine under project Ф3-19 is gratefully acknowledged. Yu.G.N. would like to thank G. E. Grechnev for the stimulating discussion on iron-chalcogenide superconductors, V. Grinenko and K. Nenkov for technical assistance. Yu.G.N. acknowledges partial support of Alexander von Humboldt Foundation in the frame of a research group linkage program. A.N.V. acknowledges support of the Ministry of Education and Science of the Russian Federation in the frames of Increase Competitiveness Program of NUST «MISiS» (№ K2-2014-036) and Russian Foundation for Basic Research (№ 14-02-92002). G.F. acknowledges support of the German Federal Ministry of Education and Research within the project ERA.Net RUS Plus: No146-MAGNES financed by the EU 7[th] FP, grant no 609556.

---

[3] As we mentioned in the introduction, the Fermi energy of FeSe is comparable to the value of the SC gap(s). This put a question about applicability of BTK and similar existing model(s) for extracting a SC gap. However, due to lack of corresponding theory, we have applied this model and, as it is seen from Fig. 5 (inset), the BTK fit is almost perfect. Anyway, such situation must be analyzed theoretically to be sure that, at least, the BTK model can be used, even in the case of $E_F \sim \Delta$.